\documentclass [12 pt]{article}
\def\be{\begin{equation}}
\def\eea{\end{eqnarray}}
\def\bea{\begin{eqnarray}}
\def\ee{\end{equation}}

\author{F. Kheirandish$^{1}$ \footnote{fardin$_{-}$kh@phys.ui.ac.ir} and M.
Amooshahi$^{1}$ \footnote{amooshahi@sci.ui.ac.ir}
\\ $^{1}$ {\small Department of Physics, University of Isfahan,}
\\ {\small Hezar Jarib Ave., Isfahan, Iran.}}
\title{Electromagnetic field quantization in a linear polarizable and magnetizable medium}
\begin{document}
\maketitle

\begin{abstract}
\noindent By modeling a linear polarizable and magnetizable medium
(magneto-dielectric) with two quantum fields, namely E and M,
electromagnetic field is quantized in such a medium consistently
and systematically. A Hamiltonian is proposed from which, using
the Heisenberg equations, Maxwell and constitutive equations of
the medium are obtained. For a homogeneous medium, the equation
of motion of the quantum vector potential, $\vec{A}$, is derived
and solved analytically. Two coupling functions which describe
the electromagnetic properties of the medium are introduced. Four
examples are considered showing the features and the
applicability of the model to both absorptive and nonabsorptive
magneto-dielectrics.


\end{abstract}
\section{Introduction}
In a homogeneous and nondispersive medium, the photon is
associated with only the transverse part of the electromagnetic
field. In contrast in an inhomogeneous nondispersive medium, the
transverse and the longitudinal degrees of freedom are coupled.
In this case the quantization of the electromagnetic field can be
accomplished by employing a generalized gauge that is,
$\vec{\nabla}\cdot(\varepsilon(\vec{r})\vec{A})=0$, where
$\varepsilon(\vec{r})$, is the space dependent dielectric
function [1,2]. Using the gauge
$\sum_{i,j=1}^3\frac{\partial}{\partial
x_i}(\varepsilon_{ij}(\vec{r})\vec{A}_j)=0$, the generalization
of this quantization to the case of an anisotropic nondispersive
medium is straightforward [3].

The quantization in a dispersive and absorptive dielectric
represents one of the most interesting problems in quantum
optics, since it gives a rigorous test of our understanding of the
interaction of light with matter. The dissipative nature of a
medium is an immediate consequence of its dispersive character
and vice versa according to the Kramers-Kronig relations. This
means that the validity of the electromagnetic field quantization
in a nondissipative but dispersive media, is restricted to a
range of frequencies for which the imaginary part of the
dielectric function is negligible. Otherwise, there will be
inconsistencies in electromagnetic field quantization process.

 In the scheme of Lenac for
dispersive and nonabsorptive dielectric media, by starting from
fundamental equations of motion, the medium is described by a
dielectric function $\varepsilon(\vec{r},\omega)$ without any
restriction on its spatial behavior [4]. In this scheme, it is
assumed that there are no losses in the system, so the dielectric
function is real for the whole space. The procedure is based on
an expansion of the total field in terms of the coupled
eigen-modes, orthogonality relations are derived and equal-time
commutation relations are discussed.

 Huttner and Barnett have
presented a canonical quantization for electromagnetic field
inside an absorptive dielectric [5]. In their model, the medium is
represented by a collection of interacting matter fields and the
absorptive character of the medium is described by interaction of
the matter fields with a reservoir containing a continuum of
Klein-Gordon fields. In this model, eigen-operators for the
coupled systems are calculated and electromagnetic field is
expressed in terms of these operators. Also the dielectric
function is derived and is shown to satisfy the Kramers-Kronig
relations.

 Gruner and
Welsch presented a quantization method of the radiation field
inside a dispersive and absorptive linear dielectric starting
from the phenomenological Maxwell equations, where the properties
of the dielectric are described by a permitivity consistent with
the Kramers-Kronig relations [6]. An expansion of the field
operators is performed which is based on the Green function of the
classical Maxwell equations and preserves the equal-time
canonical commutation relations.

 Suttorp and Wubs in the
framework of the damped polarization model, have quantized the
electromagnetic field in an absorptive medium with spatial
dependence of its parameters [7]. They have solved the equations
of motion of the dielectric polarization and the electromagnetic
field by means of the Laplace transformation for both positive
and negative times. The operators that diagonalize the
Hamiltonian are found as linear combinations of canonical
variables with coefficients depending on the electric
susceptibility and the dielectric Green function. Also the time
dependence of the electromagnetic field and the dielectric
polarization are determined.

 The macroscopic description of a
quantum damped harmonic oscillator with frequency $ \omega_0 $ is
represented in terms of the Langevin equation [8,9,10]:
\begin{equation}\label{I2}
\ddot{\vec{x}}+\int_0^\infty
dt'\mu(t-t')\dot{\vec{x}}(t')-\omega_0^2\int_0^\infty
dt'\nu(t-t')\vec{x}(t')=F_N(t).
\end{equation}
The coupling with the heat bath in the microscopic theory
corresponds to two types of forces in the macroscopic equation of
motion of a damped harmonic oscillator. The forces of the first
type are obtained from some memory functions $\mu$ and $\nu$. The
second type force is the noise force $\vec{F}_N(t)$. These two
types of forces have a fluctuation-dissipation connection and
both are required for a consistent description of a dissipative
quantum system.

Matloob has quantized the macroscopic electromagnetic field in a
linear isotropic permeable dielectric medium by quantizing the
Langevin equation and associating a damped quantum harmonic
oscillator with each mode of the radiation field [10]. There are
some other approaches to quantizing the electromagnetic field and
 interested reader is referred to [11-21].

 In this paper by modeling a polarizable and magnetizable
medium with two massless quantum fields, namely "E and M quantum
fields", we come to a completely systematic and consistent method
for quantizing electromagnetic field in such a medium. We propose
a general method in which the medium is included in quantization
process. In fact, the coupling of electromagnetic field with the
medium in a microscopic level, is modeled macroscopically by
replacing the medium with "E and M quantum fields ". The
underlying mechanism of interaction between electromagnetic field
and the medium in microscopic theory may be much more complicated
than that implied by this model which simply replace the medium
with a collection of harmonic oscillators. The present model is a
generalization of Caldeira-Leggett model where for dissipative
quantum systems, they model the environment by a collection of
harmonic oscillators [22,23]. In the Caldeira-Leggett model, the
environment's Hamiltonian is
\begin{equation}\label{Id2.1}
H_B=\sum_{n}[\frac{p_n^2}{2m_n}+\frac{1}{2}m_n\omega_n^2x_n^2],
\end{equation}
where $ m_n , x_n , p_n$ and $\omega _n$ are mass, position,
momentum and frequency of $n$th oscillator respectively. It
should be noted that the properties of the environment may in
some cases be determined on the basis of a microscopic model
which is not necessarily the Caldeira-Leggett model. As an example
we mention an Ohmic resistor which as a linear electric element
should be well described by a Hamiltonian of the form
(\ref{Id2.1}). On the other hand the underlying mechanism leading
to dissipation in a resistor may be more complicated than that
implied by the model of a collection of harmonic oscillators.

The main feature of the present approach is that "E and M quantum
fields" are to describe the electric and magnetic properties of
the medium macroscopically. It is clear (at least in the present
work, based on the results), that this assumption is an effective
one which can be compared with the phenomenological approaches to
the problem.

 In the Green function method and schemes applied in references
 \cite{[6]}, \cite{[10]}, \cite{[13]}, \cite{[14]}, \cite{[24]}
 the noise current and polarization densities are due to
interaction of electromagnetic field with the medium. Since the
kind of this interaction is not defined, the explicit forms of
the noise current and polarization densities are not known. It is
clear that the dissipative character of the medium depends on
 electric and magnetic susceptibilities. Since the
presence of the noise quantum fields are necessary for quantum
dissipative systems, there should be a dependence between the
strengths of the noise densities and the strengths of electric
and magnetic susceptibilities. Specially, when the medium becomes
a non dissipative one, i.e., when the imaginary part of the
Fourier transform of dielectric function tends to zero, the
strengths of
 the noise densities should clearly tend to zero. In the Green
function method and schemes applied in
\cite{[6]}\cite{[10]},\cite{[13]},\cite{[14]},\cite{[24]}, the
relation between the strengths of the noise densities and the
strengths of electric and magnetic susceptibilities is not clear.
Therefore it is not clear how these noises tend to zero when the
medium becomes a non dissipative medium.

The constitutive equations of the medium which relate electric
 and magnetic polarization densities to electric and magnetic fields respectively,
 should be treated as a consequence of interaction between electromagnetic field and
 the medium. But in the mentioned methods, the constitutive equations can not be
 derived from Heisenberg equations of motion. In the present approach, the coupling of
electromagnetic field with the medium is known and explicit forms
of noise densities are given. Therefore the relation between the
strengths of noise densities and the strengths of electric and
magnetic susceptibilities is clear. Also, in this approach, the
constitutive equations can be derived from the Heisenberg
equations of motion. The key role is played by what we have
introduced as coupling functions, which couple the
electromagnetic field to "E and M quantum fields". The electric
and magnetic susceptibilities are defined in terms of the coupling
functions. Also, the noise polarization densities are described
in terms of the coupling functions and the creation and
annihilation operators of the "E and M quantum fields". The
coupling functions are common factors in the noise densities and
the electric and magnetic susceptibilities.
 It can be shown that for a non dissipative medium, the noise densities vanish
  as expected.

In the damped polarization model the polarizability property of
the dielectric, is described by a quantum field Y and the
dissipative property of the medium is described by introducing a
heat bath which is independent of the field Y. The heat bath is a
continuum of Klein - Gordon fields with a continuous frequency
range. The heat bath interacts with the medium in a suitable way.
In this model, the magnetic property of the medium is not
included [5,7].

 In the present approach, the magnetic property of the
medium is included and the polarizability and the dissipative
properties of the medium are described only in terms of a single
quantum field (E quantum field). Also, the electric
susceptibility is defined in terms of a coupling function which
couples the "E quantum field" to electromagnetic field. Finally,
the electric polarization field of the medium is defined in terms
of the coupling function and the creation and
annihilation operators of the "E quantum field".\\
\section{ Quantum dynamics}
Quantum electrodynamics in a linear polarizable and magnetizable
 medium can be accomplished by modeling
 the medium with two independent quantum fields which
interact with the electromagnetic field. One of these quantum
fields namely " E quantum field", describes the polarizability
character of the medium and interacts with the displacement field
$\vec{D}$ through a minimal coupling term. The other quantum
field, namely " M quantum field", describes magnetizability
character of the medium and interacts with magnetic field through
a dipole interaction term. The Heisenberg equations for
eletromagnetic field (system) and " E and M quantum fields"
(environment), lead to both Maxwell and constitutive equations.
The constitutive equations relate the electric and magnetic
polarization densities to the macroscopic electric and magnetic
fields, respectively.

 The vector potential of the electromagnetic field in Coulomb gauge can be expanded in
terms of the plane waves as
\begin{equation} \label{d1}
\vec{A}(\vec{r},t)=\int d^3 \vec{k} \sum_{\lambda=1}^2
\sqrt{\frac{\hbar}
{2(2\pi)^3\varepsilon_0\omega_{\vec{k}}}}[a_{\vec{k}\lambda}(t)e^{i\vec{k}\cdot\vec{r}}+
a_{\vec{k}
\lambda}^\dag(t)e^{-i\vec{k}\cdot\vec{r}}]\vec{e}_{\vec{k}\lambda},
\end{equation}
where $\omega_{\vec{k}}=c|\vec{k}|$ and $\varepsilon_0 $ is the
permitivity of the vacuum. The unit vectors
$\vec{e}_{\vec{k}\lambda}, \hspace{00.40 cm} (\lambda=1,2)$ are
 polarization vectors and satisfy
\begin{eqnarray}\label{d1.5}
\vec{e}_{\vec{k}\lambda}\cdot\vec{e}_{\vec{k}\lambda'}&=&\delta_{\lambda\lambda'},\nonumber\\
 \vec{e}_{\vec{k}\lambda}\cdot\vec{k}&=&0.
\end{eqnarray}
These recent relations, guarantee that the vector potential
(\ref{d1}), satisfies the Coulomb gauge $
\vec{\nabla}\cdot\vec{A}(\vec{r},t)=0 $.

 Operators $a_{\vec{k}\lambda}(t)$ and $a_{\vec{k}\lambda}^\dag(t)
$ are annihilation and creation operators of electromagnetic
field and satisfy the following equal-time commutation rules
\begin{equation}\label{d2}
[a_{\vec{k}\lambda}(t),a_{\vec{k'}\lambda'}^\dag(t)]=
\delta(\vec{k}-\vec{k'})\delta_{\lambda\lambda'}.
\end{equation}
The conjugate canonical momentum density of the electromagnetic
field $\vec{\pi}_F(\vec{r},t)$ and also the displacement vector
operator $\vec{D}(\vec{r},t)$ are by definition
\begin{equation}\label{d3}
\vec{\pi}_F(\vec{r},t)=- \vec{D}(\vec{r},t)=i\varepsilon_0\int
d^3\vec{k} \sum_{\lambda=1}^2
\sqrt{\frac{\hbar\omega_{\vec{k}}}{2(2\pi)^3\varepsilon_0}}
[a_{\vec{k}\lambda}^\dag(t)e^{-i\vec{k}\cdot\vec{r}}-a_{\vec{k}\lambda}(t)
e^{i\vec{k}\cdot\vec{r}}]\vec{e}_{\vec{k}\lambda}.
\end{equation}
From this definition it is obvious that $\nabla\cdot \vec{D}=0$,
which is the Gauss law in the absence of external charges. The
commutation relations (\ref{d2}) lead to the commutation
relations between the components of the vector potential
$\vec{A}$ and the displacement vector operator $\vec{D}$ as
\begin{equation}\label{d3.1}
[A_l(\vec{r},t),-D_j(\vec{r'},t)]=[A_l(\vec{r},t),\pi_j(\vec{r'},t)]=
\imath\hbar\delta_{lj}^\bot(\vec{r}-\vec{r'}),
\end{equation}
where $\delta_{lj}^\bot(\vec{r}-\vec{r'})=\frac{1}{(2\pi)^3}\int
d^3\vec{k}e^{i\vec{k}\cdot(\vec{r}-\vec{r'})}(\delta_{lj}-\frac{k_l
k_j }{|\vec{k}|^2})$, is the transverse delta function
with the following properties: \\
\\
 1. Let us define the transverse and longitudinal components of an arbitrary vector
 field $\vec{F}(\vec{r},t)$, as
\begin{eqnarray}\label{d3.3}
&& \vec{F}^\bot(\vec{r},t)=\vec{F}(\vec{r},t)-\int d^3r'
\nabla'\cdot\vec{F}(\vec{r'},t)\vec{\nabla} G(\vec{r},\vec{r'}), \\
&&\vec{F}^\|(\vec{r},t)=\int d^3 r'
\nabla'\cdot\vec{F}(\vec{r'},t)\vec{\nabla} G(\vec{r},\vec{r'}),
\end{eqnarray}
respectively. Wherein $G(\vec{r},\vec{r'})$ is the Green function
\begin{equation}\label{d3.4}
G(\vec{r},\vec{r'})=-\frac{1}{4\pi|\vec{r}-\vec{r'}|}.
\end{equation}
Then as the first property, one can easily show that
\begin{equation}\label{d3.5}
F_i^\bot(\vec{r},t)=\sum_{j=1}^3\int
d^3r'\delta_{ij}^\bot(\vec{r}-\vec{r'})F_j(\vec{r'}).
\end{equation}
2. The second property is the transversality of
$\delta_{lj}^\bot(\vec{r}-\vec{r'})$
\begin{equation}\label{d3.6}
\sum_{l=1}^3\frac{\partial}{\partial
x_l}\delta_{lj}^\bot(\vec{r}-\vec{r'})=0,\hspace{1.00 cm}j=1,2,3.
\end{equation}\\
The Hamiltonian of the electromagnetic field inside a
magneto-dielectric medium, can be written as
\begin{eqnarray}\label{d4.5}
H_F(t)&=&\int d^3r [\frac{ \vec{D}^2(\vec{r},t)}{2\varepsilon_0}+
\frac{(\nabla\times\vec{A})^2(\vec{r},t)}{2\mu_0}],\nonumber\\
&=&\sum_{\lambda=1}^2\int d^3\vec{k}\hbar
\omega_{\vec{k}}a_{\vec{k}\lambda}^\dag(t) a_{\vec{k}\lambda}(t),
\end{eqnarray}
where $\mu_0$ is the magnetic permitivity of the vacuum and we
have applied the normal ordering to operators $
a_{\vec{k}\lambda}^\dag(t)$ and $a_{\vec{k}\lambda}(t)$.

 Now as mentioned, we model the medium by
two quantum fields, $E$ and $M$, which describe the electric and
magnetic properties of the medium macroscopically. Therefore, the
Hamiltonian of the medium can be written as
\begin{eqnarray}\label{d4.6}
H_d&=&H_e(t)+H_m(t), \nonumber\\
 H_e(t)&=&\sum_{\nu=1}^3\int d^3\vec{q}\int d^3\vec{k}
\hbar\omega_{\vec{k}}
d_{\nu}^\dag(\vec{k},\vec{q},t)d_{\nu}(\vec{k},\vec{q},t)
,\nonumber\\
H_m(t)&=&\sum_{\nu=1}^3\int d^3\vec{q}\int d^3\vec{k}
\hbar\omega_{\vec{k}}
b_{\nu}^\dag(\vec{k},\vec{q},t)b_{\nu}(\vec{k},\vec{q},t).\nonumber\\
&&
\end{eqnarray}
where $\omega_{\vec{k}}$ is the dispersion relation of the medium
and $ H_e $ and $ H_m $ are the Hamiltonians of the " E and M
quantum fields", respectively. The operators
$d_{\nu}(\vec{k},\vec{q},t)$, $d_{\nu}^\dag(\vec{k},\vec{q},t)$,
$b_{\nu}(\vec{k},\vec{q},t)$ and
$b_{\nu}^\dag(\vec{k},\vec{q},t)$, are annihilation and creation
operators of the " E and M quantum fields", respectively. We
impose the following equal-time commutation relations on these
operators
\begin{eqnarray}\label{d4.7}
&&[d_{\nu}(\vec{k},\vec{q},t) ,
d_{\nu'}^\dag(\vec{k'},\vec{q'},t)]=
\delta_{\nu\nu'}\delta(\vec{k}-\vec{k}')\delta(\vec{q}-\vec{q}'),\nonumber\\
&&[b_{\nu}(\vec{k},\vec{q},t),b_{\nu'}^\dag(\vec{k'},\vec{q'},t)]=
\delta_{\nu\nu'}\delta(\vec{k}-\vec{k}')\delta(\vec{q}-\vec{q}').
\end{eqnarray}
Now let us define the electric polarization density operator of
the medium as
\begin{eqnarray}\label{d4.72}
&&\vec{P}(\vec{r},t)=\sum_{\nu=1}^3 \int
\frac{d^3\vec{q}}{\sqrt{(2\pi)^3}} \int
d^3\vec{k}[f(\omega_{\vec{k}},\vec{r})d_{\nu}(\vec{k},\vec{q},t)e^{i\vec{q}\cdot\vec{r}}+H.C.]
\vec{v}_{\nu}(\vec{q}),\nonumber\\
&&
\end{eqnarray}
where for convenience H.C. stands for Hermitian Conjugation and
\begin{eqnarray}\label{d5.1}
\vec{v}_{\nu}(\vec{q})&=&\vec{e}_{\nu\vec{q}},\hspace{1.2cm}\mbox{for}\hspace{0.5cm}
\nu=1,2,\nonumber\\
\vec{v}_{3}(\vec{q})&=&\hat{q}=
\frac{\vec{q}}{|\vec{q}|},\hspace{0.5cm}\mbox{for}\hspace{0.5cm}
\nu=3.\nonumber\\
\end{eqnarray}
 In a similar manner we define the magnetic polarization density
operator of the medium as
\begin{eqnarray}\label{d4.73}
&& \vec{M}(\vec{r},t)=i\sum_{\nu=1}^3 \int
\frac{d^3\vec{q}}{\sqrt{(2\pi)^3}}\int
d^3\vec{k}[g(\omega_{\vec{k}},\vec{r})b_{\nu}(\vec{k},\vec{q},t)e^{i\vec{q}\cdot\vec{r}}-H.C.]
\vec{s}_{\nu}(\vec{q}),\nonumber\\
&&\vec{s}_{\nu}(\vec{q})=\hat{q}\times\vec{e}_{\nu\vec{q}}
,\hspace{0.5cm}\mbox{for}\hspace{1cm}\nu=1,2,\nonumber\\
&&\vec{s}_{3}(\vec{q})=\hat{q},\hspace{1.6cm}\mbox{for}\hspace{1cm}\nu=3.
\end{eqnarray}
In the polarization densities (\ref{d4.72}) and (\ref{d4.73}), the
functions
 $ f(\omega_{\vec{k}},\vec{r})$ and $g(\omega_{\vec{k}},\vec{r})$ are the coupling
 functions between the electromagnetic field and " E and M
quantum fields". The coupling functions are position dependent
(independent) for a inhomogeneous (homogeneous)
magneto-dielectrics.

 The polarization densities $\vec{P}$ and $\vec{M}$,
 are defined on the basis of the following physical assumptions:\\
 \\
 1- These densities should be Hermitian operators.\\
 2- For a linear magneto-dielectric medium, $\vec{P}$
 and $\vec{M}$, should be a linear combination of creation and
 annihilation operators of the medium. For a non-linear medium,
 $\vec{P}$ and $\vec{M}$, may not have a linear expansion in terms
 of the creation and annihilation operators of the medium.\\
 3- Polarization densities should depend on the macroscopic
properties of the medium.

Macroscopic properties of the medium will be reflected in
electric and magnetic susceptibilities. It is clear from
relations (\ref{d4.72}) and (\ref{d4.73}) that the polarization
densities $\vec{P}$ and $\vec{M}$ depend on coupling functions
$f(\omega_{k},\vec{r})$, $g(\omega_{k},\vec{r})$ and dispersion
relation ($\omega_{\vec{k}}$) of the medium. In the following we
show that electric and magnetic susceptibilities are dependent on
the coupling functions and dispersion relation.

 Now let us propose the total Hamiltonian as
\begin{eqnarray}\label{d4.55}
\tilde{H}(t)&=&\int d^3r \left\{\frac{[
\vec{D}(\vec{r},t)-\vec{P}(\vec{r},t)]^2}{2\varepsilon_0}+
\frac{(\nabla\times\vec{A})^2(\vec{r},t)}{2\mu_0}
-\nabla\times\vec{A}(\vec{r},t)\cdot\vec{M}(\vec{r},t)\right\}\nonumber\\
&+&H_e+H_m.\nonumber\\
\end{eqnarray}
Using (\ref{d3.1}) and (\ref{d3.5}), we can obtain the Heisenberg
equations for $\vec{A}$ and $\vec{D}$
\begin{equation}\label{d6}
\frac{\partial\vec{A}(\vec{r},t)}{\partial
t}=\frac{\imath}{\hbar}[\tilde{H},\vec{A}(\vec{r},t)]=
-\frac{\vec{D}(\vec{r},t)-\vec{P}^\bot(\vec{r},t)}{\varepsilon_0},
\end{equation}
\begin{equation}\label{d6.1}
\frac{\partial\vec{D}(\vec{r},t)}{\partial
t}=\frac{\imath}{\hbar}[\tilde{H},\vec{D}(\vec{r},t)]=
\frac{\nabla\times\nabla\times\vec{A}(\vec{r},t)}{\mu_0}-\nabla\times\vec{M}^\bot(\vec{r},t),
\end{equation}
where $\vec{P}^\bot$ and $\vec{M}^\bot$ are transverse components
of $\vec{P}$ and $\vec{M}$ respectively. The transverse electric
field $ \vec{E}^\bot $, magnetic induction $\vec{B}$ and magnetic
field $\vec{H}$ are
 \begin{equation}\label{d7}
 \vec{E}^\bot=-\frac{\partial\vec{A}}{\partial t},\hspace{1.00
 cm}\vec{B}=\nabla\times\vec{A},\hspace{1.00
 cm}\vec{H}=\frac{\vec{B}}{\mu_0}-\vec{M},
 \end{equation}
  so (\ref{d6}) and (\ref{d6.1}) can be rewritten as
\begin{equation} \label{d8}
\vec{D}=\varepsilon_0 \vec{E}^\bot+\vec{P}^\bot,
\end{equation}
\begin{equation}\label{d8.1}
\frac{\partial \vec{D}}{\partial t}=\nabla\times\vec{H}^\bot.
\end{equation}
In absence of an external charge density, we have
$\vec{D}^\|=\varepsilon_0\vec{E}^\|+\vec{P}^\|=0$, and we can
define the longitudinal component of the electric field as
$\vec{E}^\|=-\frac{\vec{P}^\|}{\varepsilon_0}$.

 Combining (\ref{d6}) and (\ref{d6.1}), we find
\begin{equation}\label{d9}
-\nabla^2\vec{A}+\frac{1}{c^2}\frac{\partial^2\vec{A}}{\partial
t^2}=\mu_0\frac{\partial\vec{P}^\bot}{\partial
t}+\mu_0\nabla\times\vec{M}^\bot.
\end{equation}
By using the commutation relations (\ref{d4.7}), the Heisenberg
equations for operators $ d_{\nu}(\vec{k},\vec{q},t)$ and $
b_{\nu}(\vec{k},\vec{q},t)$ can be obtained as
\begin{eqnarray}\label{d10}
\dot{d}_{\nu}(\vec{k},\vec{q},t)&=&
\frac{\imath}{\hbar}[\tilde{H},d_{\nu}(\vec{k},\vec{q},t)]\nonumber\\
&=&-\imath\omega_{\vec{k}}d_{\nu}(\vec{k},\vec{q},t)+\frac{\imath}{\hbar\sqrt{(2\pi)^3}}
\int d^3r'f^*(\omega_{\vec{k}},\vec{r'})
\vec{E}(\vec{r'},t)e^{-i\vec{q}\cdot\vec{r'}}\cdot\vec{v}_{\nu}(\vec{q}),\nonumber\\
&&
\end{eqnarray}
and
\begin{eqnarray}\label{d11}
\dot{b}_{\nu}(\vec{k},\vec{q},t)&=&
\frac{\imath}{\hbar}[\tilde{H},b_{\nu}(\vec{k},\vec{q},t)]\nonumber\\
&=-&\imath\omega_{\vec{k}}b_{\nu}(\vec{k},\vec{q},t)
+\frac{1}{\hbar\sqrt{(2\pi)^3}} \int
d^3r'g^*(\omega_{\vec{k}},\vec{r'})e^{-i\vec{q}\cdot\vec{r'}}\vec{B}(\vec{r'},t)\cdot
\vec{s}_{\nu}(\vec{q}),\nonumber\\
\end{eqnarray}
 respectively. These equations have the following formal solutions
\begin{eqnarray}\label{d11.1}
&&{d}_{\nu}(\vec{k},\vec{q},t)=
d_{\nu}(\vec{k},\vec{q},0)e^{-\imath\omega_{\vec{k}}t}+\nonumber\\
&&\frac{\imath}{\hbar\sqrt{(2\pi)^3}}\int_0^t
dt'e^{-\imath\omega_{\vec{k}}(t-t')} \int
d^3r'f^*(\omega_{\vec{k}},\vec{r'})e^{-i\vec{q}\cdot\vec{r'}}
\vec{E}(\vec{r'},t')\cdot\vec{v}_{\nu}(\vec{q}),\nonumber\\
\end{eqnarray}
\begin{eqnarray}\label{d11.2}
&&{b}_{\nu}(\vec{k},\vec{q},t)=
b_{\nu}(\vec{k},\vec{q},0)e^{-\imath\omega_{\vec{k}}t}+\nonumber\\
&&\frac{1}{\hbar\sqrt{(2\pi)^3}}\int_0^t
dt'e^{-\imath\omega_{\vec{k}}(t-t')} \int
d^3r'g^*(\omega_{\vec{k}},\vec{r'})e^{-i\vec{q}\cdot\vec{r'}}\vec{B}
(\vec{r'},t')\cdot\vec{s}_{\nu}(\vec{q}).\nonumber\\
\end{eqnarray}
 By substituting (\ref{d11.1}) in (\ref{d4.72}), we find the polarization operator

\begin{equation}\label{d12}
\vec{P}(\vec{r},t)=\vec{P}_N(\vec{r},t)+\varepsilon_0\int_0^{|t|}
d t' \chi_e(\vec{r},|t|-t')\vec{E}(\vec{r},\pm t'),
\end{equation}
where $\vec{E}=-\frac{\partial\vec{A}}{\partial
t}-\frac{\vec{P}^\|}{\varepsilon_0}$ is the total electric field.
The upper (lower) sign, corresponds to $t>0$ ($ t<0 $)
respectively.

The memory function
 \begin{eqnarray}\label{d12.5}
&&\chi_e(\vec{r},t)=\frac{8\pi}{\hbar\varepsilon_0}\int_0^\infty
d |\vec{k}|
|\vec{k}|^2|f(\omega_{\vec{k}},\vec{r})|^2\sin\omega_{\vec{k}} t \hspace{1.50cm}t>0\nonumber\\
&&\chi_e(\vec{r},t)=0\hspace{7.00cm}t\leq0
\end{eqnarray}
is called the electric susceptibility of the magneto-dielectric
which is defined in terms of dispersion relation
$\omega_{\vec{k}}$ and the coupling function $
f(\omega_{\vec{k}},\vec{r})$. The operator $\vec{P}_N(\vec{r},t) $
in (\ref{d12}) is the noise electric polarization density
\begin{eqnarray}\label{d12.6}
\vec{P}_N(\vec{r},t)=\sum_{\nu=1}^3\int\frac{d^3\vec{q}}{\sqrt{(2\pi)^3}}\int
d^3\vec{k}[f(\omega_{\vec{k}},\vec{r})d_{\nu}(\vec{k},\vec{q},0)
e^{-\imath\omega_{\vec{k}}t+i\vec{q}.\vec{r}}+H.C.]\vec{v}_{\nu}(\vec{q}).\nonumber\\
\end{eqnarray}
By using the equation (\ref{d12.5}), we can obtain the following
important relations in frequency domain
\begin{equation}\label{d12.7}
Im[\underline{\chi}_e(\vec{r},\omega)]=\frac{4\pi^2}{3\hbar\varepsilon_0}\frac{d
|\vec{k}|^3 }{d\omega}|f(\vec{r},\omega)|^2,
\end{equation}
\begin{equation}\label{d12.8}
Re[\underline{\chi}_e(\vec{r},\omega)]=\frac{8\pi}{\hbar\varepsilon_0}\int_0^\infty
d |\vec{k}|
|\vec{k}|^2|f(\omega_{\vec{k}},\vec{r})|^2\frac{\omega_{\vec{k}}}{\omega_{\vec{k}}^2-\omega^2},
\end{equation}

where
\begin{equation}\label{d12.9}
\underline{\chi}_e(\vec{r},\omega)=\int_0^\infty d t
\chi_e(\vec{r},t)e^{\imath\omega t}.
\end{equation}
 A feature of the present approach is its flexibility to
choosing an appropriate dispersion relation $\omega(|\vec{k}|)$,
such that $\frac{d \omega }{d|\vec{k}|}>0$. When a dispersion
relation is given, then knowing the susceptibility, we can obtain
the corresponding coupling function from (\ref{d12.7}) easily.
The sign of the left hand side of the relation (\ref{d12.7})
should be positive because it is the imaginary part of the
electric susceptibility in the frequency domain and this
imaginary part is also connected to losses in the medium, so it
is necessarily positive, otherwise
$\underline{\chi}_e(\vec{r},\omega)$ should be discarded, as it
would be unphysical. Therefore for consistency one must choose a
dispersion relation which is a strictly increasing function of $
|\vec{k}|$. Here we have chosen the simplest one, i.e. a linear
dispersion relation. It is remarkable to note that for a given
susceptibility $\underline{\chi}_e(\vec{r},\omega)$, by using
(\ref{d4.7}) and the definition of $ \vec{P}_N$ in (\ref{d12.6}),
one can show that for any choice of $\omega(|\vec{k}|)$ and $
f(\omega,\vec{r})$ satisfying (\ref{d12.7}), the following
commutation relations
\begin{equation}\label{d12.99}
[\underline{P}_{Ni}(\vec{r},\omega) ,
\underline{P}_{Nj}(\vec{r'},\omega')
]=\frac{\hbar\varepsilon_0}{\pi}Im[\underline{\chi}_e(\vec{r},\omega)]\delta_{ij}
\delta(\vec{r}-\vec{r'})\delta(\omega-\omega')
\end{equation}
 between the Fourier transforms of the components of the noise polarization density
 do not change. These commutation relations
are identical with those in reference \cite{[13]} and lead to the
correct commutation relations of electromagnetic field operators.
So various choices of $\omega(|\vec{k}|)$ and $ f(\omega,\vec{r})$
restricted by (\ref{d12.7}), do not affect the commutation
relations (\ref{d12.99}) and also the commutation relations
between electromagnetic field operators. This means that there
are many models with the same electric and magnetic properties
which can be taken as an environment with different dispersion
relations. In this point of view $\omega(|\vec{k}|)$ and $
f(\omega,\vec{r})$ are two free parameters of our model up to the
relation (\ref{d12.7}). As was mentioned, we choose the simplest
dispersion relation $\omega({|{k}|})=c|\vec{k}|$, where c,
velocity of light, as the proportionality coefficient, is just
for simplifying the calculations. It is clear from equation
(\ref{d12.7}) that other choices of the dispersion relation, just
lead to a redefinition of the coupling function $f(\omega
,\vec{r})$ and also more difficult mathematical expressions.

For the choice $\omega({|{k}|})=c|\vec{k}|$, the definition of the
electric susceptibility in (\ref{d12.5}), becomes
\begin{eqnarray}\label{d13}
\chi_e(\vec{r},t)&=&\frac{8\pi}{\hbar c^3
\varepsilon_0}\int_0^\infty d\omega_{\vec{k}}
\omega_{\vec{k}}^2|f(\omega_{\vec{k}}
,\vec{r})|^2\sin\omega_{\vec{k}} t,
\hspace{1cm}t>0,\nonumber\\
\chi_e(\vec{r},t)&=&0,\hspace{6.7cm} t\leq 0,
\end{eqnarray}
For a definite $\chi_e(\vec{r},t)$, which is zero for $ t\leq 0 $,
 we can obtain the corresponding coupling function
$f(\omega_{\vec{k}},\vec{r})$, in terms of $\chi_e(\vec{r},t)$, as
\begin{eqnarray}\label{d13.1}
|f(\omega_{\vec{k}} ,\vec{r})|^2&=&\frac{\hbar c^3\varepsilon_0
}{4\pi^2\omega_{\vec{k}}^2}\int_0^\infty dt\chi_e(\vec{r},t)
\sin\omega_{\vec{k}} t,\hspace{1cm}\omega_{\vec{k}} >0,\nonumber\\
|f(\omega_{\vec{k}}
,\vec{r})|^2&=&0,\hspace{5.6cm}\omega_{\vec{k}}=0.
\end{eqnarray}
Similarly, by substituting (\ref{d11.2}) in (\ref{d4.73}), we
obtain the following expression for the magnetic polarization
density $\vec{M}(\vec{r},t)$,
\begin{equation}\label{d17}
\vec{M}(\vec{r},t)=\vec{M}_N(\vec{r},t)+\frac{1}{\mu_0}\int_0^{|t|}
dt' \chi_m(\vec{r},|t|-t')\vec{B}(\vec{r},\pm t'),
\end{equation}
where $\chi_m $ is the magnetic susceptibility of the
magneto-dielectric
\begin{eqnarray}\label{d19}
\chi_m(\vec{r},t)&=&\frac{8\pi\mu_0}{\hbar c^3}\int_0^\infty
d\omega_{\vec{k}} \omega_{\vec{k}}^2|g(\omega_{\vec{k}}
,\vec{r})|^2
\sin\omega_{\vec{k}} t,\hspace{1cm}t>0,\nonumber\\
\chi_m(\vec{r},t)&=&0,\hspace{6.6cm}t\leq 0.
\end{eqnarray}
If we are given a definite $\chi_m(\vec{r},t)$, which is zero for
$ t\leq 0 $, then we can obtain the corresponding coupling
function $g(\omega_{\vec{k}},\vec{r})$ in terms of
$\chi_m(\vec{r},t)$ as
\begin{eqnarray}\label{d20}
|g(\omega_{\vec{k}} ,\vec{r})|^2&=&\frac{\hbar c^3
}{4\pi^2\mu_0\omega_{\vec{k}}^2}\int_0^\infty dt\chi_m(\vec{r},t)
\sin\omega_{\vec{k}} t,\hspace{1cm}\omega_{\vec{k}} >0,\nonumber\\
|g(\omega_{\vec{k}}
,\vec{r})|^2&=&0,\hspace{6.2cm}\omega_{\vec{k}}=0.
\end{eqnarray}
The operator $M_N(\vec{r},t)$, is the noise magnetic polarization
density
\begin{eqnarray}\label{d22}
&&\vec{M}_N(\vec{r},t)=\nonumber\\
&&i\sum_{\nu=1}^3\int\int\frac{d^3\vec{q}}{\sqrt{(2\pi)^3}}
d^3\vec{k}[g(\omega_{\vec{k}},\vec{r})b_{\nu}(\vec{k},\vec{q},0)
e^{-\imath\omega_{\vec{k}}t+i\vec{q}\cdot\vec{r}}-H.C.]
\vec{s}_{\nu}(\vec{q}).\nonumber\\
&&
\end{eqnarray}
It is remarkable to note that the constitutive equations
(\ref{d8}), (\ref{d12}) and (\ref{d17}) together with Maxwell
equations are obtained directly from the Heisenberg equations
applied to the electromagnetic field and the quantum fields $ E $
and $ M $. The explicit forms of the noise polarization densities
are given by (\ref{d12.6}) and (\ref{d22}). The coupling
functions $f$ and $g$, are common factors in the noise densities
$\vec{P}_N$ and $\vec{M}_N $ and the susceptibilities $ \chi_e$,
$\chi_m $. So it is clear that the strengths of the noise fields
are dependent on the strengths of the susceptibilities $\chi_e$
and $\chi_m $, which describe the dissipative character of a
magneto-dielectric medium.

 For a homogeneous medium, the coupling functions $f(\omega_{\vec{k}},\vec{r})$ and $
g(\omega_{\vec{k}},\vec{r})$, are position independent. In this
case, from (\ref{d13}) and (\ref{d19}), we deduce that $\chi_e$
and $\chi_m $, are also position independent. Substituting
(\ref{d12}) and (\ref{d17}) in the wave equation (\ref{d9}), we
find
\begin{eqnarray}\label{d23}
&&-\nabla^2\vec{A}+\frac{1}{c^2}\frac{\partial^2\vec{A}}{\partial
t^2}\pm \frac{1}{c^2}\frac{\partial}{\partial t}\int_0^{|t|}
dt'\chi_e(|t|-t')\frac{\partial \vec{A}}{\partial
t'}(\vec{r},\pm t')-\nonumber\\
&&\nabla\times\int_0^{|t|} dt'\chi_m(|t|-t')\nabla\times
\vec{A}(\vec{r},\pm t')
=\mu_0\frac{\partial\vec{P}^\bot_N}{\partial
t}(\vec{r},t)+\mu_0\nabla\times\vec{M}_N(\vec{r},t),\nonumber\\
&&
\end{eqnarray}
where $ c^2=\frac{1}{\varepsilon_0\mu_0} $ and the upper (lower)
sign corresponds to $t>0$ ($t<0$) respectively.

The equation (\ref{d23}) is the Langevin-Schr\"{o}dinger equation
[8] for vector the potential $ \vec{A}$, wherein, the explicit
form of the noise current density is known.

The quantum Langevin equation can be considered as the basis of the
macroscopic description of a quantum particle coupled to an
environment or a heat bath \cite{[8]},\cite{[9]}.\\
\section{ Solution of the Heisenberg equations}
In this section, we solve the Heisenberg equations for the vector
potential. Let us denote the Fourier transform of the vector
potential $\vec{A}(\vec{r},t)$ by $
\underline{\vec{A}}(\vec{q},t)$, so
\begin{equation}\label{d24}
\vec{A}(\vec{r},t)=\frac{1}{\sqrt{(2\pi)^3}}\int
d^3\vec{q}\underline{\vec{A}}(\vec{q},t)e^{i\vec{q}\cdot\vec{r}},
\end{equation}
from (\ref{d1}), it is clear that
\begin{equation}\label{d26}
\underline{\vec{A}}(\vec{q},t)=\sum_{\lambda=1}^2
\sqrt{\frac{\hbar} {2
\varepsilon_0\omega_{\vec{q}}}}[a_{\vec{q}\lambda}(t)\vec{e}_{\vec{q}\lambda}+a_{-\vec{q}
\lambda}^\dag(t)\vec{e}_{-\vec{q}\lambda}].
\end{equation}
The wave equation (\ref{d23}) can be written in terms of $
\underline{\vec{A}}(\vec{q},t)$ as
\begin{eqnarray}\label{d28}
&&\underline{\ddot{\vec{A}}}+\omega_{\vec{q}}^2\underline{\vec{A}}
\pm\frac{\partial}{\partial t}\int_0^{|t|}
dt'\chi_e(|t|-t')\dot{\underline{\vec{A}}}(\vec{q},\pm
t')-\omega_{\vec{q}}^2\int_0^{|t|}
dt'\chi_m(|t|-t')\underline{\vec{A}}(\vec{q},\pm t')\nonumber\\
&=&-\frac{\imath }{\varepsilon_0} \sum_{\lambda=1}^2 \int\frac{
d^3\vec{k}}{\sqrt{(2\pi)^3}}[\omega_{\vec{k}}f(\omega_{\vec{k}})d_{\lambda}(\vec{k},\vec{q},0)
e^{-\imath\omega_{\vec{k}}t}\vec{e}_{\vec{q}\lambda}-\omega_{\vec{k}}f^*(\omega_{\vec{k}})
d_{\lambda}^\dag(\vec{k},-\vec{q},0)e^{\imath\omega_{\vec{k}}t}\vec{e}_{-\vec{q}\lambda}]
\nonumber\\
&+&\omega_{\vec{q}}\sqrt{\frac{\mu_0}{\varepsilon_0}}\sum_{\lambda=1}^2\int
\frac{d^3\vec{k}}{\sqrt{(2\pi)^3}}[g(\omega_{\vec{k}})b_{\lambda}
(\vec{k},\vec{q},0)e^{-\imath\omega_{\vec{k}}t}\vec{e}_{\vec{q}\lambda}+
g^*(\omega_{\vec{k}})b_{\lambda}^\dag(\vec{k},-\vec{q},0)
e^{\imath\omega_{\vec{k}}t}\vec{e}_{-\vec{q}\lambda}],\nonumber\\
&&
\end{eqnarray}
where $\omega_{\vec{q}}=c|\vec{q}|$.

This equation  can be solved using the Laplace transformation
method. For any time dependent operator $g(t)$, the forward and
backward Laplace transformations are by definition
\begin{equation}\label{d28.1}
g_f(s)=\int_0^\infty dt g(t)e^{-st},
\end{equation}
 and
\begin{equation}\label{d28.2}
g_b(s)=\int_0^\infty dt g(-t)e^{-st},
\end{equation}
respectively. Let $ \tilde{\chi}_e(s)$ and $\tilde{\chi}_m(s)$ be
the Laplace transformations of $ \chi_e(t) $ and $\chi_m(t)$
respectively. Then $ \underline{\vec{A}}_f(\vec{q},s) $ and $
\underline{\vec{A}}_b(\vec{q},s) $, i.e., the forward and
backward Laplace transformation of $\vec{A}(\vec{q},t) $, can be
obtained in terms of $\tilde{\chi}_e(s)$ and $\tilde{\chi}_m(s)$
as
\begin{eqnarray}\label{d29}
&&\underline{\vec{A}}_{f,b}(\vec{q},s)=\frac{s+s\tilde{\chi}_e(s)}{s^2+
\omega_{\vec{q}}^2+s^2\tilde{\chi}_e(s)-\omega_{\vec{q}}^2\tilde{\chi}_m(s)}
\underline{\vec{A}}(\vec{q},0)\nonumber\\
&&\pm\frac{1}{s^2+\omega_{\vec{q}}^2+s^2\tilde{\chi}_e(s)-\omega_{\vec{q}}^2
\tilde{\chi}_m(s)}\dot{\underline{\vec{A}}}(\vec{q},0)\nonumber\\
&&-\frac{\imath}{\varepsilon_0}\sum_{\lambda=1}^2\int
d^3\vec{k}\{\frac{\omega_{\vec{k}}f(\omega_{\vec{k}})d_\lambda(\vec{k},\vec{q},0)}
{(s\pm\imath\omega_{\vec{k}})[s^2+\omega_{\vec{q}}^2+s^2\tilde{\chi}_e(s)-\omega_{\vec{q}}^2
\tilde{\chi}_m(s)]}\vec{e}_{\vec{q}\lambda}\nonumber\\
&&-\frac{\omega_{\vec{k}}f^*(\omega_{\vec{k}})d_\lambda^\dag(\vec{k},-\vec{q},0)}
{(s\mp\imath\omega_{\vec{k}})[s^2+\omega_{\vec{q}}^2+s^2\tilde{\chi}_e(s)-\omega_{\vec{q}}^2
\tilde{\chi}_m(s)]}\vec{e}_{-\vec{q}\lambda}\}\nonumber\\
&&+\sqrt{\frac{\mu_0}{\varepsilon_0}}\omega_{\vec{q}}\sum_{\lambda=1}^2\int
d^3\vec{k}\{\frac{g(\omega_{\vec{k}})b_\lambda(\vec{k},\vec{q},0)}{(s\pm\imath
\omega_{\vec{k}})[s^2+\omega_{\vec{q}}^2+s^2\tilde{\chi}_e(s)-\omega_{\vec{q}}^2
\tilde{\chi}_m(s)]}\vec{e}_{\vec{q}\lambda}\nonumber\\
&&+\frac{g^*(\omega_{\vec{k}})b_\lambda^\dag(\vec{k},-\vec{q},0)}{(s\mp\imath\omega_{\vec{k}})
[s^2+\omega_{\vec{q}}^2+s^2\tilde{\chi}_e(s)-\omega_{\vec{q}}^2\tilde{\chi}_m(s)]}
\vec{e}_{-\vec{q}\lambda}\}
\end{eqnarray}
where the upper (lower) sign corresponds to $
\underline{\vec{A}}_f(\vec{q},s)$
($\underline{\vec{A}}_b(\vec{q},s)$) respectively. Now taking the
inverse Laplace transformation of $
\underline{\vec{A}}_f(\vec{q},s) $ and $
\underline{\vec{A}}_b(\vec{q},s)$, we obtain a complete solution
for $\vec{A}(\vec{r},t)$
\begin{eqnarray}\label{d35.1}
\vec{A}(\vec{r},t)&=&\sum_{\lambda=1}^2\int
d^3\vec{q}\sqrt{\frac{\hbar}{2(2\pi)^3\varepsilon_0\omega_{\vec{q}}}}[Z_{\pm}
(\omega_{\vec{q}},t)e^{\imath\vec{q}\cdot\vec{r}}a_{\vec{q}\lambda}(0)+H.C.]
\vec{e}_{\vec{q}\lambda}\nonumber\\
&\pm&\frac{1}{\varepsilon_0}\sum_{\lambda=1}^2\int
\frac{d^3\vec{q}}{\sqrt{(2\pi)^3}}\int
d^3\vec{k}[\zeta_{\pm}(\omega_{\vec{k}},\omega_{\vec{q}},t)d_\lambda(\vec{k},\vec{q},0)
e^{\imath\vec{q}\cdot\vec{r}}+H.C.]\vec{e}_{\vec{q}\lambda}\nonumber\\
&+&\sqrt{\frac{\mu_0}{\varepsilon_0}}\sum_{\lambda=1}^2\int
\frac{d^3\vec{q}\omega_{\vec{q}}}{\sqrt{(2\pi)^3}}\int
d^3\vec{k}[\eta_{\pm}(\omega_{\vec{k}},\omega_{\vec{q}},t)b_\lambda(\vec{k},\vec{q},0)
e^{\imath\vec{q}\cdot\vec{r}}+H.C.]\vec{e}_{\vec{q}\lambda},\nonumber\\
&&
\end{eqnarray}
where the upper (lower) sign corresponds to $ t>0 (t<0)$. The
functions $Z_{+}(\omega_{\vec{q}},t)$,
$Z_{-}(\omega_{\vec{q}},-t)$, $\zeta_{+}(\omega_{\vec{k}}$,
$\omega_{\vec{q}},t)$, $\zeta_{-}(\omega_{\vec{k}}$,
$\omega_{\vec{q}},-t)$,
 $\eta_{+}(\omega_{\vec{k}}$, $\omega_{\vec{q}},t)$,
 $\eta_{-}(\omega_{\vec{k}},\omega_{\vec{q}},-t)$ are given by
\begin{eqnarray}\label{d36}
&&Z_{\pm}(\omega_{\vec{q}},\pm
t)=L^{-1}\left\{\frac{[s+s\tilde{\chi}_e(s)\mp\imath\omega_{\vec{q}}]
}{s^2+\omega_{\vec{q}}^2+s^2\tilde{\chi}_e(s)-\omega_{\vec{q}}^2\tilde{\chi}_m(s)}\right\}
\nonumber\\
&&\zeta_{\pm}(\omega_{\vec{k}},\omega_{\vec{q}},\pm
t)=f(\omega_{\vec{k}})L^{-1}\left\{\frac{s
}{(s\pm\imath\omega_{\vec{k}})[s^2+\omega_{\vec{q}}^2+s^2\tilde{\chi}_e(s)-
\omega_{\vec{q}}^2\tilde{\chi}_m(s)]}\right\}\nonumber\\
&&\eta_{\pm}(\omega_{\vec{k}},\omega_{\vec{q}},\pm
t)=g(\omega_{\vec{k}})L^{-1}\left\{\frac{1
}{(s\pm\imath\omega_{\vec{k}})[s^2+\omega_{\vec{q}}^2+s^2\tilde{\chi}_e(s)-
\omega_{\vec{q}}^2\tilde{\chi}_m(s)]}\right\},\nonumber\\
&&
\end{eqnarray}
for $ t>0 $ and $ L^{-1}{f(s)}$ is the inverse Laplace
transformation of function $f(s)$. The transverse component of
the electric field can be obtained from $ \vec{E}^\bot
=-\frac{\partial\vec{A}}{\partial t} $. Having the vector
potential $\vec{A}$, the transverse component of electric
polarization density ($ \vec{P}^\bot$) and also magnetic
polarization density ($\vec{M}$), can be obtained easily from
relations (\ref{d12}) and (\ref{d17}).

 Taking the Laplace transformation of the
constitutive equation (\ref{d12}), the longitudinal component of
the electric field can be written as
\begin{eqnarray}\label{d37}
\vec{E}^\|(\vec{r},t)&=&-\frac{\vec{P}^\|}{\varepsilon_0}\nonumber\\
&=&-\frac{1}{\varepsilon_0}\int \frac{d^3\vec{q}}{\sqrt{(2\pi)^3}}
\int
d^3\vec{k}[Q_\pm(\omega_{\vec{k}},t)f(\omega_{\vec{k}})d_3(\vec{k},\vec{q},0)
e^{i\vec{q}\cdot \vec{r}}+H.C.]\hat{q},\nonumber\\
&&
\end{eqnarray}
where $ Q_{+}(\omega_{\vec{k}},t)$, $Q_{-}(\omega_{\vec{k}},-t)$
are given by
\begin{equation}\label{d38}
Q_\pm(\omega_{\vec{k}},\pm t)= L^{-1}\left\{\frac{1
}{(1+\tilde{\chi}_e(s))(s\pm i\omega_{\vec{k}})}\right\},
\end{equation}
for $ t>0 $. In the following we consider some important examples.\\
\\
\textbf{Example 1:}\\
Let $f(\omega_{\vec{k}})=g(\omega_{\vec{k}})=0 $, then from
(\ref{d13}) and (\ref{d19}), we have $ \chi_e(t)=\chi_m(t)=0 $,
and from (\ref{d36}), we find
\begin{equation}\label{d39}
Z_+(\omega_{\vec{q}},t)=Z_-(\omega_{\vec{q}},t)=e^{-i\omega_{\vec{q}}t},\hspace{1.50
cm } \eta_\pm=\zeta_\pm=0,
\end{equation}
therefore in this limiting case, quantization of electromagnetic
field reduce to the usual quantization in the vacuum as expected.\\
\\
\textbf{Example 2 :}\\
Take $\chi_e(t)$ and $\chi_m(t)$ as follows
\begin{displaymath}\label{d40}
\chi_e(t)=\left\{\begin{array}{ll}
\frac{\chi_e^0}{\triangle} & 0<t<\triangle,\\
0 & \textrm{otherwise},
\end{array}\right.
\end{displaymath}
\begin{displaymath}\label{d41}
\chi_m(t)=\left\{\begin{array}{ll}
\frac{\chi_m^0}{(\chi_m^0+1)\triangle} & 0<t<\triangle\\
0 & \textrm{otherwise,}
\end{array} \right.
\end{displaymath}
\begin{equation}\label{d42}
\end{equation}
where $\chi_e^0$, $\chi_m^0$ and $\triangle $ are some positive
constants, using (\ref{d13.1}) and (\ref{d20}), we find the
corresponding coupling functions as
\begin{eqnarray}\label{d42.1}
|f(\omega_{\vec{k}})|^2&=&\frac{\hbar
c^3\varepsilon_0\chi_e^0}{4\pi^2
\omega_{\vec{k}}^2}\frac{\sin^2\frac{\omega_{\vec{k}}\triangle}{2}}
{\frac{\omega_{\vec{k}}\triangle}{2}},\nonumber\\
|g(\omega_{\vec{k}})|^2&=&\frac{\hbar
c^3\chi_m^0}{4\pi^2\mu_0(\chi_m^0+1)
\omega_{\vec{k}}^2}\frac{\sin^2\frac{\omega_{\vec{k}}\triangle}{2}}
{\frac{\omega_{\vec{k}}\triangle}{2}},\nonumber\\
\end{eqnarray}
and from (\ref{d12}) and (\ref{d17}) we have
\begin{eqnarray}\label{d42.2}
&&\vec{P}(\vec{r},t)=\vec{P}_N(\vec{r},t)+\frac{\varepsilon_0\chi_e^0}{\triangle}
\int_{|t|-\triangle}^{|t|} d t'\vec{E}(\vec{r},\pm t'),\nonumber\\
&&\vec{M}(\vec{r},t)=\vec{M}_N(\vec{r},t)+\frac{\chi_m^0}{\mu_0(\chi_m^0+1)
\triangle}\int_{|t|-\triangle}^{|t|} d t'\vec{B}(\vec{r},\pm t'),
\end{eqnarray}
where $\vec{P}_N(\vec{r},t)$ and $\vec{M}_N(\vec{r},t)$ are the
noise polarization densities (\ref{d12.6}) and (\ref{d22})
corresponding to the coupling functions obtained in (\ref{d42.1}).

In the limit $\triangle \rightarrow0 $, the coupling functions
(\ref{d42.1}) and the noise polarization densities, tend to zero,
and the relations (\ref{d42.2}) are reduced to
\begin{eqnarray}\label{d42.3}
\vec{P}(\vec{r},t)&=&\varepsilon_0\chi_e^0\vec{E}(\vec{r},t),\nonumber\\
\vec{M}(\vec{r},t)&=&\frac{\chi_m^0}{\mu_0(\chi_m^0+1)}\vec{B}(\vec{r},t).\nonumber\\
\end{eqnarray}
In this limit, the electric field and the polarization densities
are purely transverse and
\begin{eqnarray}\label{d42.4}
&& Z_-(\omega_{\vec{q}},t)=Z_+(\omega_{\vec{q}},t)=
\cos\tilde{\omega}_{\vec{q}}t-i\sqrt{\frac{1+\chi_m^0}{1+\chi_e^0}}
\sin\tilde{\omega}_{\vec{q}}t,\nonumber\\
&&\tilde{\omega}_{\vec{q}}=\frac{\omega_{\vec{q}}}{\sqrt{(1+\chi_e^0)(1+\chi_m^0)}},\nonumber\\
&&\eta_\pm(\omega_{\vec{q}},t)=\zeta_\pm(\omega_{\vec{q}},t)=0.
\end{eqnarray}
Also the electromagnetic energy inside the dielectric is
\begin{eqnarray}\label{d45}
\int[\frac{1}{2}\vec{E}\cdot\vec{D}+\frac{1}{2}\vec{H}\cdot\vec{B}]d^3r&=&
\int d^3\vec{q}[\frac{1}{2\varepsilon_0(1+\chi_e^0)}
\underline{\vec{D}}(\vec{q}.0)\cdot
\underline{\vec{D}}^\dag(\vec{q}.0)\nonumber\\
&+&\frac{\varepsilon_0\omega_{\vec{q}}^2}{2(1+\chi_m^0)}\underline{\vec{A}}(\vec{q},0)\cdot
\underline{\vec{A}}^\dag(\vec{q},0)]
\end{eqnarray}
where $\underline{\vec{D}}$ is the Fourier transform of the
displacement field. The energy given by \ref{d45}) is a constant
of motion contrary to the vacuum expression
$\int[\frac{1}{2}\varepsilon_0\vec{E}^2+\frac{\vec{B}^2}{2\mu_0}]d^3r$,
which is not clearly a constant of motion. This example shows that
this
model can be applied to a nondispersive magneto-dielectric medium.\\
\\
\textbf{Example 3:}
 Let $\chi_e(t)=\beta u(t)$ and $\chi_m(t)=0$, where $ u(t) $ is
 the step function
 \begin{displaymath}
u(t)=\left\{\begin{array}{ll}
1 & t>0\\
0 & t\leq 0
\end{array}\right.
\end{displaymath}
\begin{equation}\label{d46}
\end{equation}
and $ \beta $ is a positive constant, then using (\ref{d13.1}) and
(\ref{d20}), we find
\begin{equation}\label{d47}
|f(\omega_{\vec{k}})|^2=\frac{\hbar
c^3\varepsilon_0\beta}{4\pi^2\omega_{\vec{k}}^3},\hspace{2.00
cm}g(\omega_{\vec{k}})=0,
\end{equation}
and accordingly we can rewrite (\ref{d28}) as
\begin{eqnarray}\label{d48}
\ddot{\underline{\vec{A}}}+\omega_{\vec{q}}^2 \underline{\vec{A}}
+\beta\dot{\underline{\vec{A}}}&=& -\imath\sqrt{\frac{\hbar
c^3\beta}{4\pi^2 \varepsilon_0}}
\int\frac{d^3\vec{k}}{\sqrt{(2\pi)^3\omega_{\vec{k}}}}\sum_{\lambda=1}^2[d_{\lambda}
(\vec{k},\vec{q},0)e^{-\imath\omega_{\vec{k}}t}\vec{e}_{\vec{q}\lambda}\nonumber\\
&-&d_{\lambda}^\dag(\vec{k},-\vec{q},0)e^{\imath
\omega_{\vec{k}}t}\vec{e}_{-\vec{q}\lambda}],
\end{eqnarray}
which has a dissipative term proportional to the first time
derivative of dynamical variable $\underline{\vec{A}}$.  From
(\ref{d36}), one can obtain
\begin{eqnarray}\label{d48.1}
&&Z_\pm(\omega_{\vec{q}},t)=e^{\mp\frac{\beta}{2}t}[\pm\frac{\beta}
{2\Omega_{\vec{q}}}\sin\Omega_{\vec{q}}t+
\cos\Omega_{\vec{q}}t-\frac{i\omega_{\vec{q}}}{\Omega_{\vec{q}}}
\sin\Omega_{\vec{q}}t],\nonumber\\
&&\zeta_\pm(\omega_{\vec{k}},\omega_{\vec{q}},t)=\sqrt{\frac{\hbar
c^3\beta\varepsilon_0}{4\pi^2\omega_{\vec{k}}^3}}\left\{\mp\frac{i\omega_{\vec{k}}
e^{-i\omega_{\vec{k}}t}}{\omega_{\vec{q}}^2-\omega_{\vec{k}}^2\mp
i\beta\omega_{\vec{k}}}+\right.\nonumber\\
&&e^{\mp\frac{\beta}{2}t}[\frac{(-\frac{\beta}{2}+i\Omega_{\vec{q}})e^{\pm
i\Omega_{\vec{q}}t}}{2i\Omega_{\vec{q}}(-\frac{\beta}{2}+i\Omega_{\vec{q}}\pm
i\omega_{\vec{k}}
)}\left.+\frac{(\frac{\beta}{2}+i\Omega_{\vec{q}})e^{\mp
i\Omega_{\vec{q}}t}}{2i\Omega_{\vec{q}}(-\frac{\beta}{2}-i\Omega_{\vec{q}}\pm
i\omega_{\vec{k}} )}]\right\},
\end{eqnarray}
where
$\Omega_{\vec{q}}=\sqrt{\omega_{\vec{q}}^2-\frac{\beta^2}{4}}$.
The asymptotic solution of $\vec{A}(\vec{r},t)$ in large-time
limit is
\begin{eqnarray}\label{d49}
\vec{A}(\vec{r},t)=\mp\imath\sqrt{\frac{\hbar c^3\beta}{4\pi^2
\varepsilon_0 }}\sum_{\lambda=1}^2
\int\frac{d^3\vec{q}}{\sqrt{(2\pi)^3}}
\int\frac{d^3\vec{k}}{\sqrt{\omega_{\vec{k}}}}\left[\frac{d_{\lambda}
(\vec{k},\vec{q},0)e^{-\imath\omega_{\vec{k}}t+i\vec{q}\cdot\vec{r}}}
{\omega_{\vec{q}}^2-\omega_{\vec{k}}^2\mp\imath\beta
\omega_{\vec{k}}}-H.C.\right].\nonumber\\
&&
\end{eqnarray}
Using (\ref{d37}), it is easy to show that the longitudinal
component of the electric field in the limit $t\rightarrow
\pm\infty$ is
\begin{eqnarray}\label{d49.1}
\vec{E}^\|(\vec{r},t)&=&-\frac{\vec{P}^\|}{\varepsilon_0}\nonumber\\
&=&\pm i\sqrt{ \frac{\hbar
c^3\beta}{4\pi^2\varepsilon_0}}\int\frac{d^3\vec{q}}{\sqrt{(2\pi)^3}}
\int
\frac{d^3\vec{k}}{\sqrt{\omega_{\vec{k}}}}\left[\frac{d_3(\vec{k},\vec{q},0)}{\beta\mp
i\omega_{\vec{k}} }e^{-i\omega_{\vec{k}}t+i\vec{q}\cdot
\vec{r}}-H.C.\right].\nonumber\\
&&
\end{eqnarray}
\\
\textbf{Example 4: A simple model for $ \tilde{\chi}_e(s) $ }\\
If we neglect the difference between local and macroscopic
electric field for substances with a low density, then the
classical equation of a bound atomic electron in an external
electric field is
\begin{equation}\label{d51}
\ddot{\vec{r}}+\gamma\dot{\vec{r}}+\omega_0^2\vec{r}=-\frac{e}{m}\vec{E}(t),
\end{equation}
where the influence of the magnetic force has been neglected
compared to the electric force. The parameter $\gamma$ is a
damping coefficient and the force exerted on the electron due to
atom is taken to be simply a spring force with frequency $
\omega_0 $. If $ \tilde{\vec{E}}(s) $ and $ \tilde{\vec{r}}(s) $
are the Laplace transformations of $\vec{E}(t) $ and $ \vec{r}(t)
$ respectively, then from (\ref{d51}) we can find
\begin{equation}\label{d52}
\tilde{\vec{r}}(s)
=\frac{-\frac{e}{m}\tilde{\vec{E}}(s)}{s^2+\gamma s+\omega_0^2}.
\end{equation}
Now let there be $ N $ molecules per unit volume with $ z $
electrons per molecule such that $ f_j $ electrons of any
molecule have a bound frequency $ \omega_j $ and a damping
coefficient $ \gamma_j $. The Laplace transformation of the
polarization density is
\begin{equation}\label{d53}
\tilde{\vec{P}}(s)=\frac{Ne^2}{m}\sum_j\frac{f_j}{s^2+\gamma_js+\omega_j^2}
\tilde{\vec{E}}(s).
\end{equation}
If $ \omega_j $ and $ \gamma_j $ are identical for all of
electrons, then from (\ref{d53}) we can write
  \begin{eqnarray}\label{d54}
&&\tilde{\chi}_e(s)=\frac{\omega_p^2}{s^2+\gamma
s+\omega_0^2},\hspace{1.5
cm}\omega_p^2=\frac{Ne^2z}{m\varepsilon_0},\nonumber\\
&&\chi_e(t)=\omega_p^2e^{-\frac{\gamma
t}{2}}\frac{\sin\nu_0t}{\nu_0}u(t),\hspace{0.7
cm}\nu_0^2=\omega_0^2-\frac{\gamma^2}{4},
\end{eqnarray}
where $ u(t) $ is the step function defined in (\ref{d46}). We can
obtain the coupling function $ f(\omega_{\vec{k}}) $ from
(\ref{d13.1}) as
 \begin{eqnarray}\label{d55}
 |f(\omega_{\vec{k}})|^2=\frac{\hbar c^3\varepsilon_0\omega_p^2}{16\pi^2\nu_0
 \omega_{\vec{k}}^2}\left\{\frac{\gamma}{\frac{\gamma^2}{4}+(\nu_0-\omega_{\vec{k}})^2}-
 \frac{\gamma}{\frac{\gamma^2}{4}+(\nu_0+\omega_{\vec{k}})^2}\right\}.
\end{eqnarray}
If $ \gamma=0 $, then the dielectric substance is a
nondissipative one and the coupling function takes the form
\begin{equation}\label{d58}
|f(\omega_{\vec{k}})|^2=\frac{\hbar
c^3\varepsilon_0\omega_p^2}{8\pi\nu_0^3}\delta(\nu_0-\omega_{\vec{k}}).
\end{equation}
In this case, the noise electric polarization density is nonzero
only for the resonant frequency ($\omega=\omega_0 $) of the
equation $ \ddot{\vec{r}}+\omega_0^2
\vec{r}=-\frac{e}{m}\vec{E}_0 e^{-i\omega_0 t}$. In the resonant
case, energy of electromagnetic field will be absorbed by the
medium. By using (\ref{d36}), we can obtain $
Z_{+}(\omega_{\vec{q}},t)$ and $Z_{-}(\omega_{\vec{q}},-t)$, as
\begin{equation}\label{d59}
Z_\pm(\omega_{\vec{q}},\pm t) = L^{-1}\left\{\frac{(s\mp
i\omega_{\vec{q}})(s^2+\omega_0^2)+s\omega_p^2}{s^4+s^2
(\omega_0^2+\omega_{\vec{q}}^2+\omega_p^2)
+\omega_{\vec{q}}^2\omega_0^2}\right\},
\end{equation}
 for
$ t>0 $. The equation (\ref{d59}) can be solved by calculating the
residues of the function
\begin{equation}\label{d60}
\frac{[(s\mp
i\omega_{\vec{q}})(s^2+\omega_0^2)+s\omega_p^2]e^{\pm st}
}{s^4+s^2(\omega_0^2+\omega_{\vec{q}}^2+\omega_p^2)
+\omega_{\vec{q}}^2\omega_0^2},
\end{equation}
by extending s to the domain of complex variables. Similarly
$\zeta_\pm$ can be obtained from (\ref{d36}) with $
f(\omega_{\vec{k}}) $ given by (\ref{d58}). In this case we have
\begin{eqnarray}\label{d61}
Q_\pm(\omega_{\vec{k}},t)&=&\frac{\omega_0^2-\omega_{\vec{k}}^2}{\omega_0^2+
\omega_p^2-\omega_{\vec{k}}^2}e^{-i\omega_{\vec{k}}t}\nonumber\\
&+&\frac{\omega_p^2}
{2\sqrt{\omega_0^2+\omega_p^2}}\left\{\frac{e^{\pm
i\sqrt{\omega_0^2+\omega_p^2}t}}{\sqrt{\omega_0^2+\omega_p^2}\pm\omega_{\vec{k}}}+
\frac{e^{\mp
i\sqrt{\omega_0^2+\omega_p^2}t}}{\sqrt{\omega_0^2+\omega_p^2}
\mp\omega_{\vec{k}}}\right\}.
\end{eqnarray}
The longitudinal component of the electric field can be obtained
from (\ref{d37}), using $f(\omega_{\vec{k}})$ and $
Q_\pm(\omega_{\vec{k}},t)$ given by (\ref{d58}) and (\ref{d61})
respectively.

 If $\gamma\neq 0$, the substance is of
dissipative kind and $\vec{A}(\vec{r},t)$, can be obtained from
(\ref{d35.1}) with $ f(\omega_{\vec{k}})$ given by (\ref{d55})
and $\tilde{\chi}_e(s)$ given by (\ref{d54}). In this case the
longitudinal component of the electric field can be obtained from
(\ref{d37}). This example shows that this model of quantization
of the electromagnetic field is applicable to both dissipative
and nondissipative dielectrics.
\section{Concluding remarks}
By modeling a linear, polarizable and magnetizable medium with two
 quantum fields E and M, electromagnetic field
 is quantized in the medium, consistently and systematically.
 There are many models for the environment with the same electric
  and magnetic properties. In other words, there are many dispersion
  relations and coupling functions leading to the same suceptibilities of the medium.
  Therefore one can take the simplest physical dispersion relation.
  Once a dispersion relation is defined, then for any definite magneto-dielectric medium, i.e.,
   $\chi_e(t)$ and $\chi_m(t)$ are known functions, one can find the corresponding
coupling functions $f(\omega_{\vec{k}},\vec{r})$ and
$g(\omega_{\vec{k}},\vec{r})$. The coupling functions describe the
electric and magnetic properties of the medium macroscopically.
These functions also couple electromagnetic field to the medium
through the quantum fields E and M. The explicit forms of the
noise densities are derived. Since the coupling functions are
common factors in the noise densities and susceptibilities, the
relation between the strengths of the noise densities and
 susceptibilities is clear. In this
approach, both Maxwell and constitutive equations are obtained as
Heisenberg equations of motion. In the limiting case, i.e., when
there is no medium, the approach tends to the usual quantization
of the electromagnetic field in vacuum as expected. This model of
quantization is applicable to both dispersive and nondispersive
magneto-dielectrics.

\end{document}